\documentclass[apjl]{emulateapj}

\newcommand{\degree}{\ensuremath{\mathrm{^\circ}}}

\newcommand{\fermi}{\emph{Fermi}~}

\newcommand{\fermilat}{\emph{Fermi}~LAT~}
\newcommand{\fermilatnospace}{\emph{Fermi}~LAT}

\begin{document}
\shorttitle{3FGL Galactic Demographics}
\shortauthors{Mirabal et al.}
\title{3FGL Demographics Outside the Galactic Plane 
  using Supervised Machine Learning:
Pulsar and Dark Matter Subhalo Interpretations}
\author{
  N. Mirabal,\altaffilmark{1,2} E. Charles, \altaffilmark{3} E. C. Ferrara,\altaffilmark{1}
  P. L. Gonthier,\altaffilmark{4} A. K. Harding,\altaffilmark{1} M. A. S\'anchez-Conde,\altaffilmark{5,6} \&
  D. J. Thompson\altaffilmark{1}} 
\affil{
$^1$ NASA Goddard Space Flight Center, Greenbelt, MD 20771, USA\\
  $^2$ NASA Postdoctoral Program Fellow, USA\\
  $^3$ W. W. Hansen Experimental Physics Laboratory, Kavli Institute
for Particle Astrophysics and Cosmology, Department of Physics and SLAC National Accelerator Laboratory, Stanford University, Stanford, CA 94305, USA\\
$^4$ Hope College, Department of Physics, Holland, MI 49423 USA\\
$^5$Department of Physics, Stockholm University, AlbaNova, SE--106 91 Stockholm,Sweden\\
$^6$The Oskar Klein Centre for Cosmoparticle Physics, AlbaNova, SE--106 91 
Stockholm, Sweden}

\begin{abstract} 
Nearly 1/3 of the sources listed in the Third \fermi Large Area Telescope 
(LAT) catalog (3FGL)
remain unassociated. It is possible that predicted and even
unanticipated gamma-ray source classes 
are present in these data  waiting to be discovered. 
Taking advantage of the excellent spectral capabilities achieved by the
\fermilatnospace, 
we use machine learning classifiers (Random Forest and
XGBoost) to pinpoint potentially novel source classes 
in the unassociated 3FGL sample outside the Galactic plane. 
Here we report a total of 34 high-confidence  
Galactic candidates at  $|b| \geq 5\degree$.
The currently favored standard astrophysical interpretations for these
objects are pulsars or low-luminosity globular clusters hosting
millisecond pulsars (MSPs). Yet, these objects could also be
interpreted as dark matter annihilation taking place in ultra-faint
dwarf galaxies or dark matter subhalos. Unfortunately, 
\fermilat spectra
are not sufficient to break degeneracies between the different scenarios.
Careful visual inspection of 
archival optical images reveals no obvious evidence for low-luminosity globular
clusters or ultra-faint dwarf galaxies inside 
the 95\% error ellipses. 
If these are pulsars, this would bring the total
number of MSPs 
at $|b| \geq 5\degree$ to 106, down to an energy flux  
$\approx 4.0 \times10^{-12}$ erg 
cm$^{-2}$ s$^{-1}$ between 100 MeV and 100 GeV.   
We find this number to be in excellent agreement with predictions from 
a new population synthesis of MSPs that predicts 
100--126 high-latitude 3FGL
MSPs depending on the choice of high-energy emission model.
If, however, these are dark matter substructures, 
we can place
upper limits on the number of Galactic subhalos surviving today and
on dark matter
annihilation cross sections. These limits
are beginning to approach the canonical thermal
relic cross section for dark matter particle masses below $\sim100$ GeV
in the bottom quark 
($b\bar{b}$) annihilation channel. 
\end{abstract}

\section{Introduction}
There is compelling evidence for the existence of dark matter in
the Universe.
Already in 1933, \citet{zwicky} had collected enough data to postulate
the presence of 
more mass than what could be inferred from visible galaxies.
Over the past 80 years, additional observations ranging from
the rotation curves of
spiral galaxies \citep{rubin} to large-scale structure \citep{planck,betoule}
seem to point in the same direction. One outstanding prediction of
cosmological simulations using cold dark matter particles is
that the Milky Way halo should be heavily populated with 
thousands of smaller dark matter subhalos
as a result of the hierarchical assembly process
\citep{klypin,moore,springel}.
Dark matter subhalos would include any dark matter configuration, from those
hosting the largest known dwarf spheroidal galaxies
in the Milky Way to the lightest predicted dark matter substructures with 
masses around $10^{-4} M_{\odot}$ \citep{ricotti,scott}.
Conceptually in this scenario, the bulk of the subhalo population 
is made up by small-scale dark matter substructures 
with limited or null star formation,
which would be almost impossible to detect in existing optical surveys. 

Should dark matter subhalos without major star formation episodes
exist in large numbers, one of the only ways to detect them 
might be by tracking gamma rays from dark matter annihilation.
The all-sky coverage and unprecedented sensitivity of the Large
Area Telescope (LAT), on board NASA's \fermi satellite, is enabling
the most effective search for potential dark matter subhalos to date.
The obvious place to look for subhalo
candidates is among the 3033 sources detected and characterized
in the Third {\it Fermi}--LAT catalog (3FGL) \citep{3fgl},
in particular, 
among the 1010  
sources listed as unassociated with counterparts of known
gamma-ray-producing source classes. 
It is plausible that 
after years of observations, \fermi has already detected 
dark matter subhalos. The challenge now is to locate them  \citep{charles}. 

Previous attempts to pinpoint dark matter subhalos have systematically
searched for unassociated sources
with spectra that are consistent with dark matter
annihilation \citep{buckley,nieto,belikov,zechlin,satellites,berlin,bertoni,bertoni2}.
Machine-learning
classification targeting dark matter subhalos 
has also been performed using $k$-means clustering
in the First \fermilat Catalog (1FGL)
\citep{mirabal} and Random Forest in
the Second \fermilat Catalog (2FGL) \citep{mirabal3}.
Here we present an application of machine-learning classifiers  
that aims to chart all Galactic sources outside the plane in the 3FGL. By 
specifically isolating 3FGL objects at high latitude we hope to  
reduce the search space for dark matter subhalos and unknown
source classes as much as possible.
The paper is organized as follows.
In Section 2 we introduce the spectral prescription for 
locating Galactic objects outside the plane in the 3FGL. 
In Section 3 we introduce machine-learning classifiers.
Section 4 covers training sample and variable selection. 
In Sections 5 and 6 we discuss cross validation
and prediction results. In Section 7, 
we discuss possible interpretations of the results including
a visual search for ultra-faint dwarf galaxy and globular cluster
counterparts, as well as 
direct comparison with statistics from pulsar population synthesis models.
We also place
constraints on the annihilation cross section by comparing the
number of potential subhalos to predictions from 
cosmological simulations.
Finally, in Section 8 we present our
conclusions and outlook.

\section{Spectral Prescription for the Galactic Population outside the plane}
At high Galactic latitude, 
the totality of known Galactic gamma-ray emitters
correspond to pulsars or globular clusters hosting MSPs \citep{3fgl}. 
Two additional undiscovered gamma-ray-producing source classes have been
postulated to exist: dark matter subhalos \citep{bergstrom99} and  
dwarf galaxies \citep{lake}. Dark matter annihilation is expected
to dominate the gamma-ray signal from both known Galactic
dwarf galaxies and from dark matter subhalos. 
If two dark matter particles annihilate through typical Standard Model 
channels, the decay and hadronization of these
particles would create a gamma-ray spectrum that 
extends up to the rest mass of the dark matter particle
with a sharp cutoff \citep{bergstrom,fornengo}.
The gamma-ray spectrum is given by

\begin{equation}\label{eq1}
\frac{d\Phi}{dE}(E,\Delta\Omega) = \frac{1}{4\pi}
\frac{\langle\sigma v\rangle}{2 m^2_\chi}
\frac{dN_{\gamma}}{dE} \times J(\Delta\Omega),
\end{equation}

where $\langle\sigma v\rangle$ is the
thermally averaged annihilation cross section,
$m_{\chi}$ is the dark matter particle mass, and
$\frac{dN_{\gamma}}{dE}$ is the gamma-ray yield per annihilation
and depends on the particle physics model under consideration \citep{pythia}.

The second part of the equation,
or the so-called astrophysical factor $J(\Delta\Omega)$,
corresponds to the
integration of the dark matter density squared $\rho^2(l,\Omega)$  along the
line of sight $l$
over a solid angle $\Delta\Omega$. The nearest and/or most massive
dark matter subhalos would be easiest to detect even if they host
no stars. 

Figure 1 shows the
predicted gamma-ray
spectrum for the annihilation of 30 GeV dark matter 
particles into bottom quark
($b\bar{b}$) pairs. The spectral shape
from dark matter annihilation
would be quite recognizable. Unfortunately, a dark matter
annihilation spectrum around a few GeV would not be
unique. Most known gamma-ray pulsars display nearly
identical spectra with sharp exponential cutoffs
\citep{baltz,2fpc}. To illustrate this point, 
we have overlaid in Figure 1 a typical MSP spectrum drawn 
from the Second \fermilat Pulsar Catalog (2FPC) \citep{2fpc}.

This pulsar-dark matter spectral
degeneracy has driven a lively debate
about the origin of excess GeV emission at the Galactic center 
\citep{gh,hl,abazajian,mirabal2,gordon,yuan,daylan,carlson,petrovic,calore2,brandt,bartels,lee,fermidm,oleary};
however, the spectral degeneracy permeates to other 
astrophysical settings beyond the Galactic center. Indeed,
because of this gamma-ray degeneracy, one cannot expect to 
distinguish spectrally between 
pulsars, low-luminosity globular
clusters hosting MSPs, ultra-faint dwarf 
galaxies, and dark matter subhalos.

By contrast, none of the known extragalactic sources
have exponentially curved LAT spectra, except for
the Large Magellanic Cloud (LMC) pulsar \citep{lmcpulsar}. Variability 
is also a far more likely characteristic of extragalactic objects
than of Galactic objects at high latitude. 
One can turn the spectral degeneracy between some Galactic
classes and the marked contrast with extragalactic
objects into assets by using known pulsar and
globular cluster gamma-ray spectra to help  hunt down  
undiscovered dark matter subhalos and ultra-faint dwarf galaxies in the 3FGL 
with machine-learning classifiers.
We describe such search strategy next.

\begin{figure}[!ht]
\epsscale{1.1}\plotone{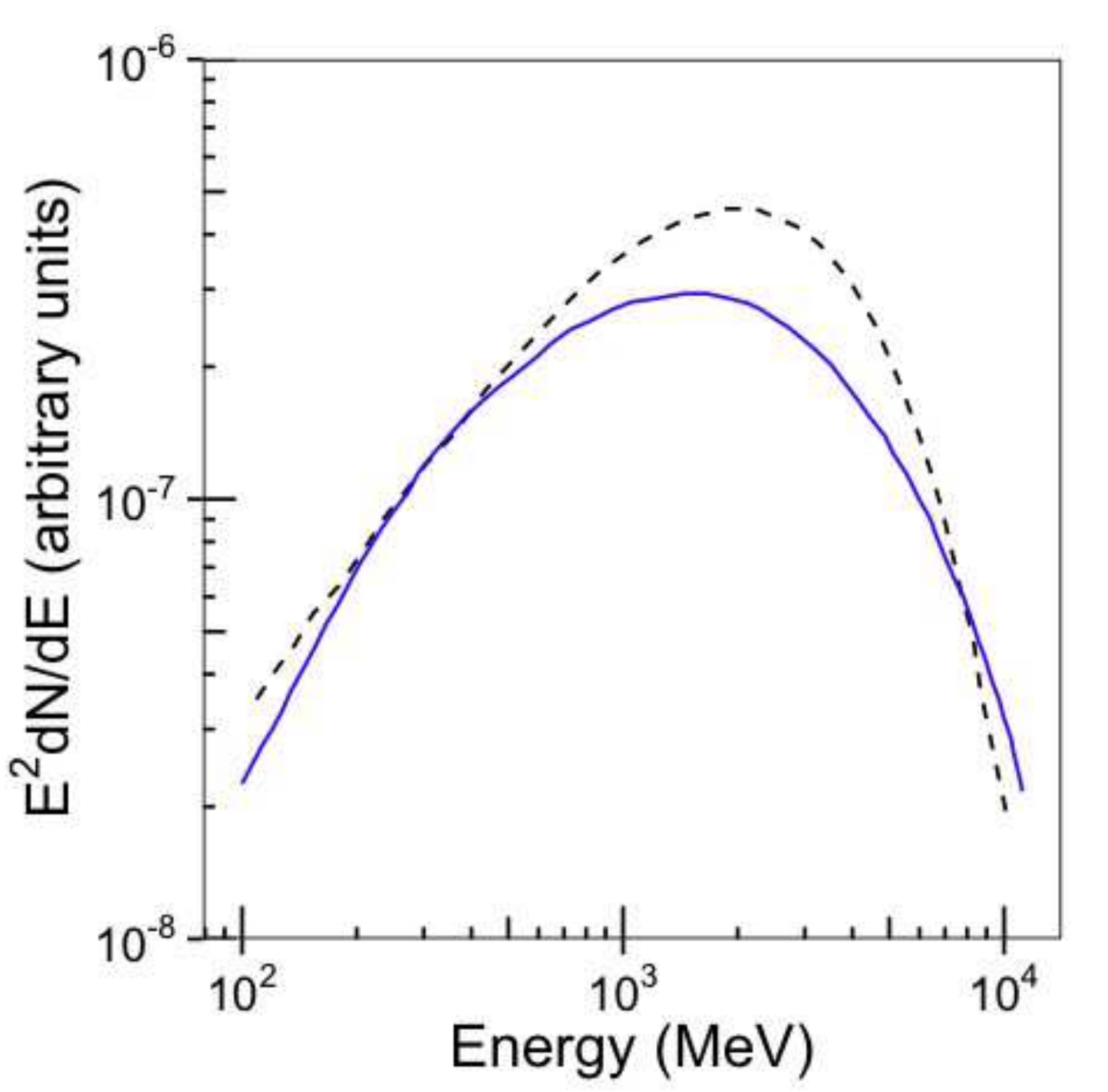}
\caption{Spectral shape from dark matter 
annihilation (blue line) of a 30 GeV particle into  
 bottom quark 
($b\bar{b}$) pairs \citep{fornengo}.
Other Standard Model annihilation channels are expected to 
produce similar spectra. Also shown
is a representative MSP
spectrum (black dashed line) from the 2FPC \citep{2fpc}.
}
\end{figure}
 
\section{Machine-Learning Classifiers}
By machine learning classifier, 
we denote any algorithm with the ability to perform accurate predictions, 
after having trained on the properties of a well-known training set of data.   
For most of this work, 
we will concentrate on supervised machine learning classifiers or sets
of models that can
input a list of variables of a data set and output
a prediction model that best describes the relationship between the
variables and known classes. For an overview of popular machine-learning
classifiers, we refer the reader to \citet{hastie}.
Among the plethora of available
machine-learning classifiers,  
we have settled on   
two highly accurate variants of classification trees, i.e.
Random Forest and XGBoost \citep{delgado,chen}.

\subsection{Random Forest}
Random Forest is an 
ensemble learning method that grows a large forest of 
randomized classification trees and aggregates their predictions made 
from a list of input variables
\citep{breiman}. Individual classification
trees are constructed by randomly sampling $k$ variables from $n$ input
variables at each node \citep{quinlan}.  
The forest selects a random sample with replacement
from the original training set using
bagging (bootstrap aggregating). Any data left out of the bootstrap sample
helps to measure the internal accuracy 
directly in the form of an out-of-bag estimates.   
To classify an object, 
each tree in the forest issues a prediction. 
The predictions from all trees for the same
object are then collected and
a class is determined through a majority decision. 
Aside from being easy to use, Random Forest provides outstanding performance
and the ability to track
 proximity matrices. 
We adopted  the
\textit{randomForest} set of routines, which implements
the original Random Forest for classification 
and regression to the R language \citep{liaw}.
The final model is based on 1000 trees, with
a total of $\sqrt{p}$ variables sampled at each split,
where $p$ is the final number of variables.

\subsection{XGBoost}
The eXtreme Gradient 
Boosting XGBoost\footnote{https://github.com/dmlc/xgboost} is
a modified version of gradient boosting \citep{friedman}.
The fundamental 
difference with Random Forest 
is that XGBoost uses boosting to reweigh
the training set sequentially \citep{quinlan}. In contrast to bagging,
boosting uses all instances at each repetition but issues a weight for
each instance in the training set. These evolving weights adjust the
learners to focus on different instances at each pass. 
One of the key problems in tree classifiers is how to find the best split
at each node. To expedite this decision,
XGBoost finds the best solution over all the
possible splits according to percentiles of variable distribution \citep{chen}. 
In all experiments, we boost trees with 
a learning rate $\eta = 0.5$ and a
maximum
number of iterations set to 5. 

\section{Training Sample and Variable Selection}
In order to construct a base training set for the Galactic population
outside the plane, we use  
143 3FGL pulsars identified by pulsations (PSRs), 15 3FGL globular
clusters (glcs),  and 24 3FGL pulsars with no pulsation
seen in LAT yet (psrs).  
The extragalactic set includes 1745 sources  
from all AGN classes in the 3FGL. It contains
3 non-blazar active galaxies (agn), 
573 blazar candidates of uncertain type (BCU,bcu), 660 BL Lacs (BLL,bll), 
484 flat-spectrum radio quasars (FSRQ,fsrq), 
5 narrow-line Seyferts 1 (NLSY1,nlsy1), 
15 radio galaxies (RDG,rdg), 3 soft spectrum radio quasars (ssrq), 
1 compact steep spectrum quasar (css), and 1 Seyfert galaxy (sey). The entire 
dataset is divided into a training set (a random sample of 2/3 of the total) and
a testing set (remaining 1/3 of the total).

Variable selection is
essential to machine-learning classifiers. Starting from a 
set of parameters or variables the algorithm must be able to classify an
object into one of a set of distinct classes. 
For any classification problem, there are certain variables 
that best capture a specific class of objects. The advantage
of using machine-learning classifiers
is that the algorithms can explore the entire variable space at once. 
Initially we started with a total of 35 3FGL variables, excluding positional,
uncertainty and descriptive variables.  
In view of the inherent difficulties in sampling gamma-ray spectra 
for any given object, we 
augmented this original
variable set with eight additional derived variables defined by
hardness ratios 
\texttt{HR$_{ij}$}= $\frac{Flux_i - Flux_j}{Flux_i + Flux_j}$ and
flux 
ratios \texttt{FR$_{ij}$} = \texttt{Flux$_{i}$}/ \texttt{Flux$_j$} between
consecutive $i,j$
bands for 0.1--0.3 GeV (Band 1), 0.3--1 GeV (Band 2), 1--3 GeV (Band 3), 
3--10 GeV (Band 4), 
and
10--100 GeV (Band 5).

Figure 2 shows the out-of-bag error as a function of the 
number of variables used. The out-of-bag estimates
tend to asymptote to a minimum at 
$n \gtrsim 8$. 
The value shown for $n = 2$ corresponds to the optimal scenario when using
only \texttt{Signif\_Curve} and \texttt{Variability\_Index} as variables.  
For each classifier, we also quantify feature importance to pinpoint
the features that best discriminate between classes. 
Table 1 ranks variable importance (from most to least important) in terms of 
the  improvement achieved from each variable or $Gain$ for all splits and
trees for the XGboost model. For comparison, 
Table 2 ranks the overall percentage decrease in accuracy rate
averaged over all trees
or $MeanDecreaseAccuracy$ measured by Random Forest \citep{liaw}. 
As can be seen, the variable rankings are not identical but there
is commonality on the top variables. The differences are an indication
of the distinct paths the classifiers follow to achieve 
a successful prediction.
After exploring different variable combinations, we find that the 
variables that most clearly differentiate Galactic and extragalactic 
populations include 
\texttt{Signif\_Curve}, \texttt{Variability\_Index}, and
\texttt{Spectral\_Index}. We also include flux ratios and
hardness ratios for five energy bands in our final models.

\begin{figure}
\epsscale{1.2}\plotone{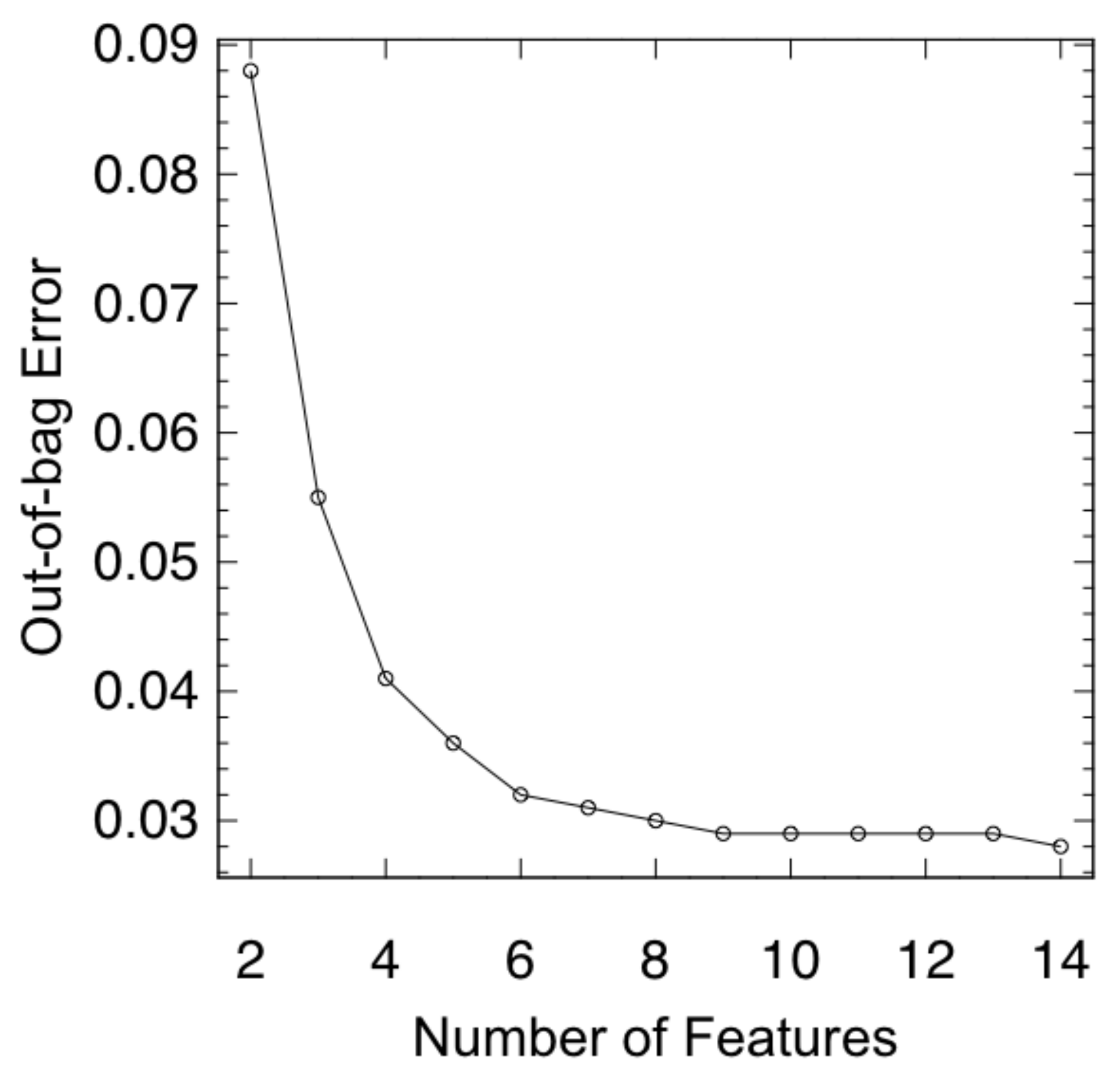}
\caption{Out-of-bag estimates as a function of the number of variables used by the
classifier.}
\end{figure}

\begin{deluxetable}{cc}
\tablecaption{3FGL variable importance according to $Gain$ in XGBoost.}

\tablehead{\colhead{Variable} & \colhead{$Gain$}} \\ 
\startdata
\texttt{Signif\_Curve}   &      0.47 \\
\texttt{Variability\_Index}  & 0.26 \\ 
\texttt{FR$_{45}$} & 0.10 \\ 
\texttt{Spectral\_Index} & 0.08 \\
\texttt{FR$_{12}$}  &  0.03 \\
\texttt{FR$_{23}$}   & 0.03 \\
\texttt{Pivot\_Energy} & 0.02 \\
\texttt{FR$_{34}$}  &  0.01 \\
\enddata

\end{deluxetable} 

\begin{deluxetable}{cc}
\tablecaption{3FGL variable importance according to $MeanDecreaseAccuracy$ in Random Forest.}
\tablenum{2}
\tablehead{\colhead{Variable} & \colhead{$MeanDecreaseAccuracy$}} \\ 
\startdata
\texttt{Variability\_Index} & 0.58\\
\texttt{FR$_{45}$} & 0.40\\
\texttt{Signif\_Curve} & 0.40\\
\texttt{Pivot\_Energy} & 0.33\\
\texttt{HR$_{45}$} & 0.30\\
\texttt{Spectral\_Index} & 0.26\\
\texttt{HR$_{23}$}  &  0.24\\
\texttt{FR$_{23}$}  &  0.22\\
\enddata
\end{deluxetable} 

\section{Cross Validation}

\begin{figure}
\epsscale{1.2}\plotone{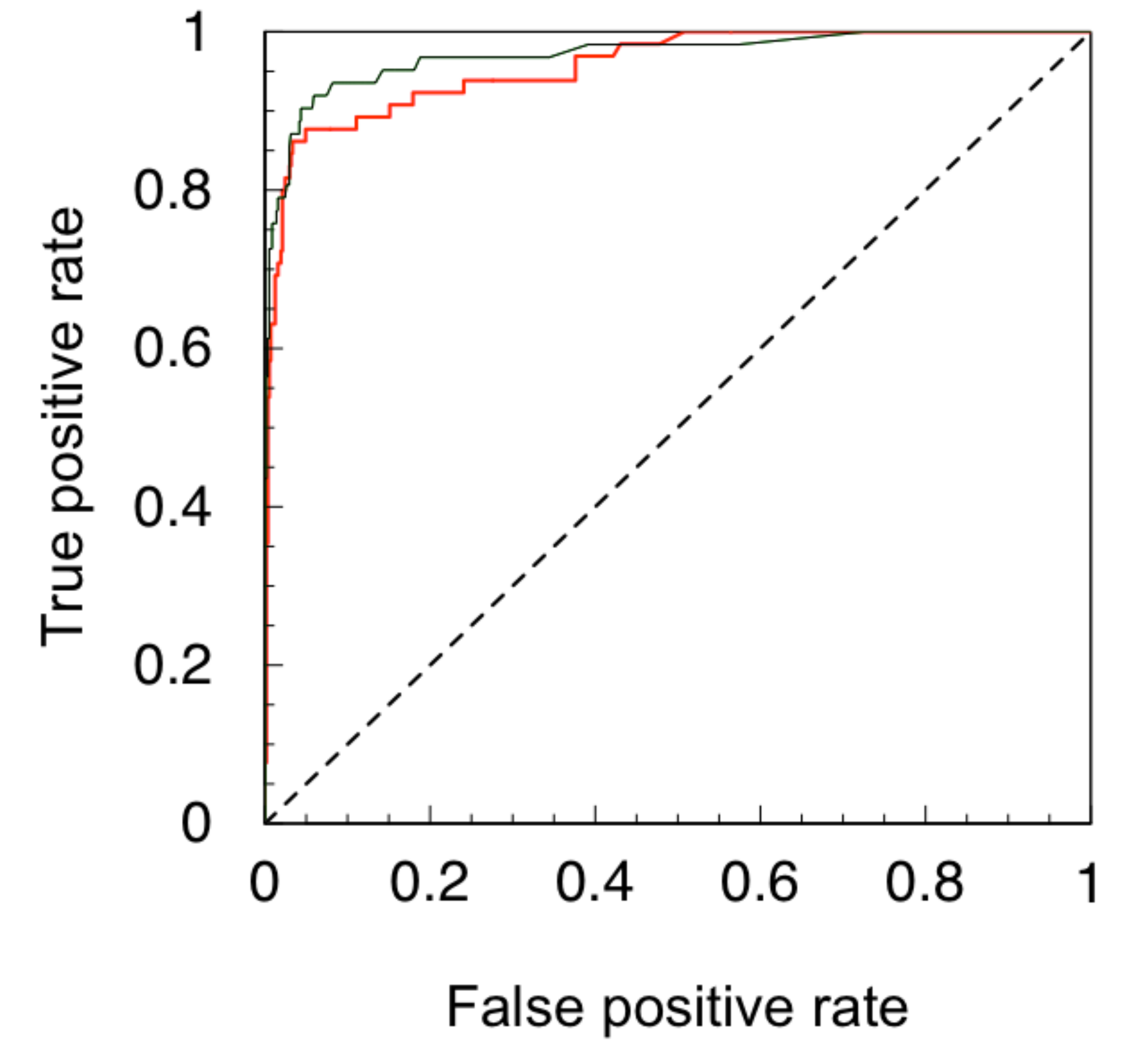}
\caption{
  The Receiver Operating Characteristic (ROC)
  curves for Random Forest(black line) and XGBoost (red line). The
  closer the curve to the upper left corner, the better the performance.
  The $45^{\circ}$ dashed line marks the region where true positive
  and false positive rates are equal.
}
\end{figure}

\begin{figure}[!ht]
\epsscale{1.1}\plotone{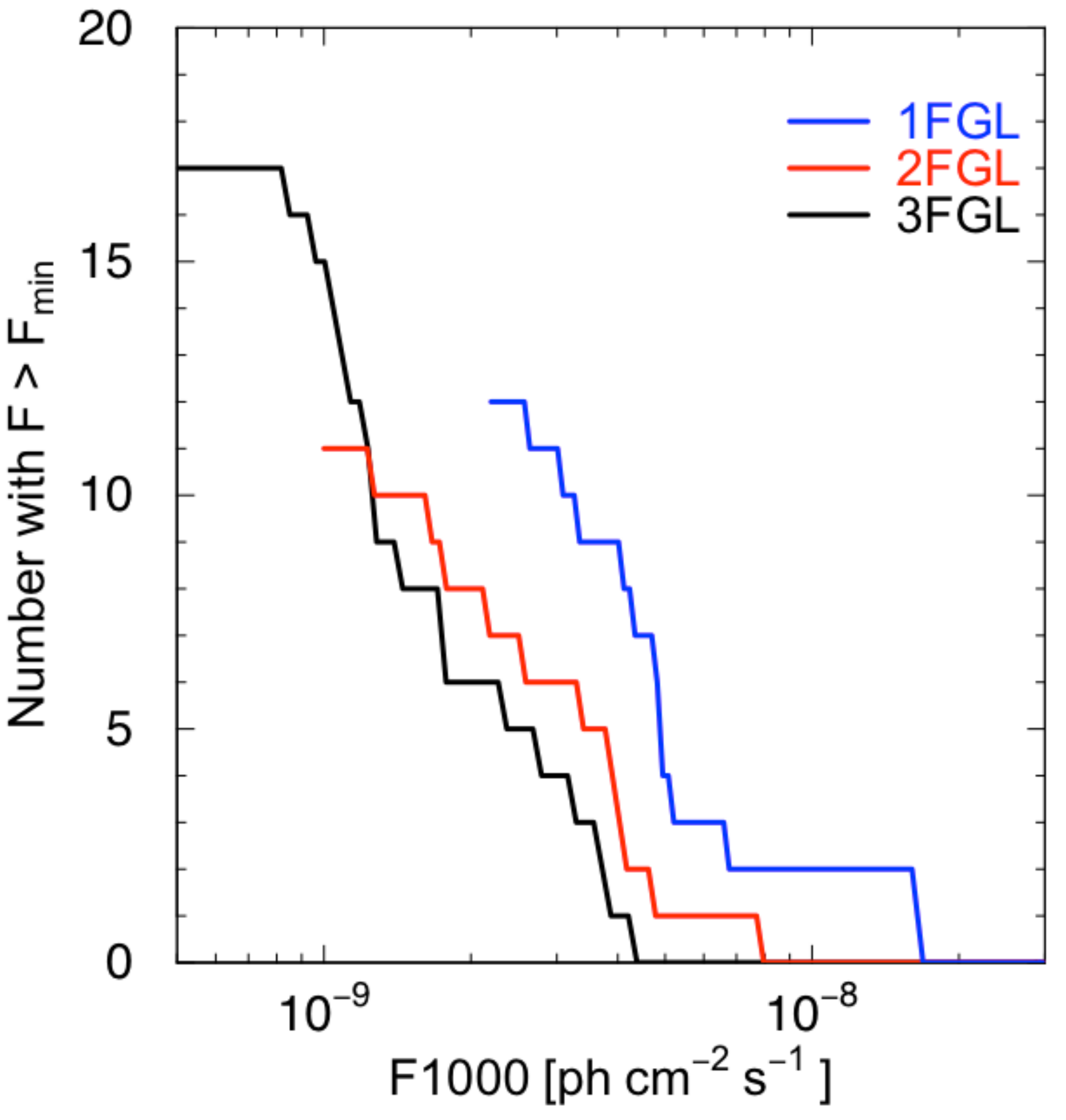}
\caption{Number of potential 3FGL Galactic candidates at $|b| \geq 20\degree$
(black).
Also shown are 1FGL (blue) and 2FGL (red) candidates
using the same machine-learning
approach. F1000 represents the photon flux for the 1--100 GeV energy range.}
\end{figure}

\begin{figure*}[!ht]
\epsscale{0.7}\plotone{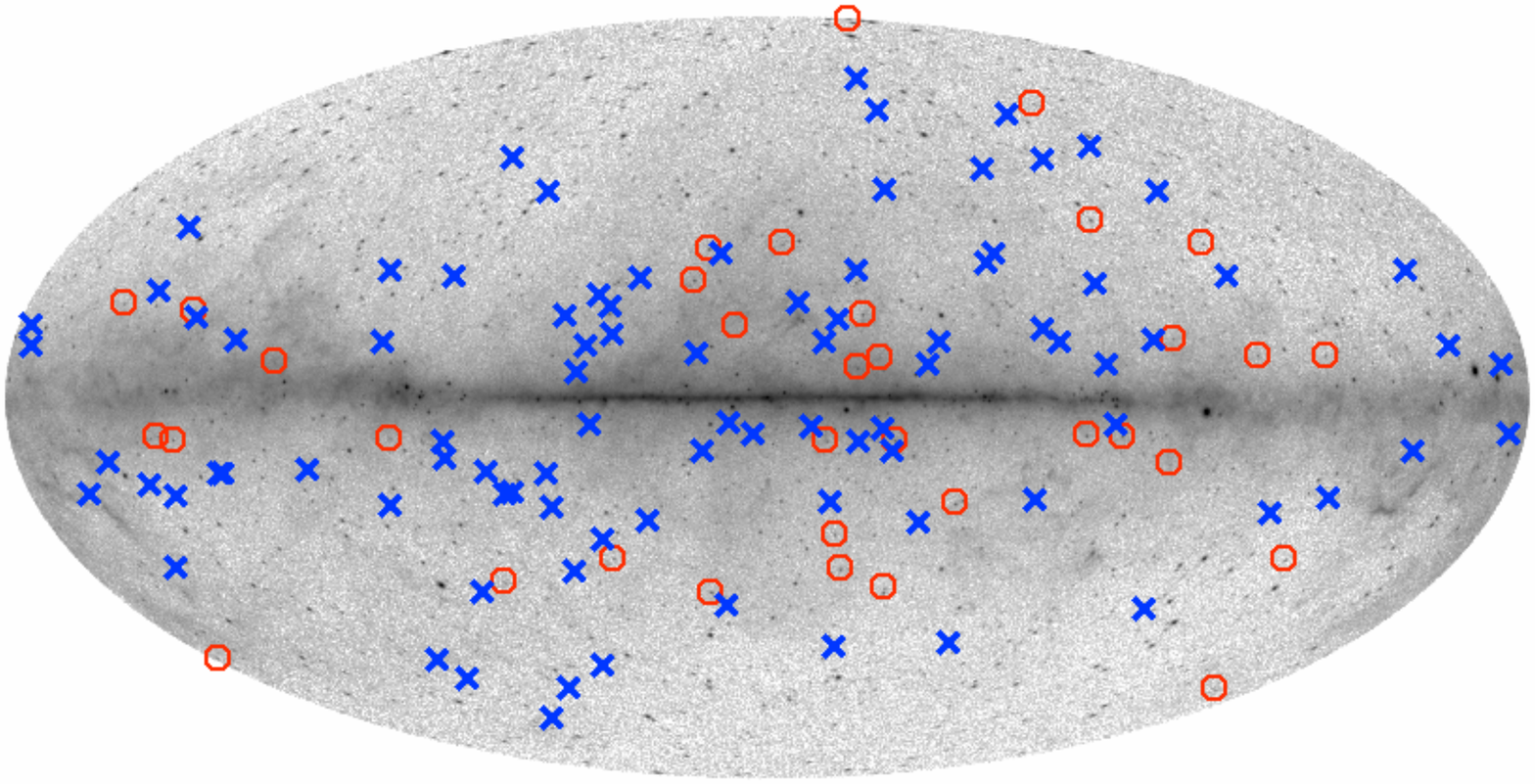}
\caption{Galactic distribution of 34 high-latitude Galactic
  candidates (red circles)
  superimposed on a smoothed \fermilat all-sky map for energies $E \geq 1$ GeV
  based on events collected during the period 2008 August 4--2015 August 4
  (Credit: \fermilat Collaboration). 
High-latitude 3FGL pulsars (blue crosses) are also plotted for comparison.
  }
\end{figure*}

During initial tests, we quickly found out that 
Random Forest suffered from the fact that
the training set was highly imbalanced, i.e.
90\% extragalactic versus 10\% Galactic
trainees. As constructed, Random Forest tends to build classification 
trees dominated by  the majority class. To alleviate this problem,  
we artificially increased the  training set
using over-sampling as implemented
in the Synthetic Minority Over-sampling Technique ($SMOTE$) \citep{chawla}.
$SMOTE$ creates a much larger synthetic minority class by replicating 
the existing sample using five nearest neighbors. With the   
$SMOTE$ approach we were able to achieve a test set
error rate of 2.8\% for Random Forest. 
Class imbalanced was not a problem for XGBoost and the training 
was done with the original training and testing
samples, reaching a test set error rate of 3.3\%.

Figure 3 shows the receiver operating characteristic (ROC) curves
for both classifiers \citep{hastie}. At various threshold values, we plot
the true positive rate (the fraction of
correctly identified) versus the false positive rate (the fraction of
candidates incorrectly
classified). As shown, Random Forest slightly outperforms
XGBoost, but only after taking into account the class imbalance. 
A perfect classifier would have a true positive rate equal
to one.

A complementary measure of accuracy  
can be accomplished by applying
Random Forest and XGBoost retroactively to unassociated sources
listed in the 1FGL and 2FGL. Since a number of 
{\it Fermi} discoveries have been made
since those early catalogs were first released, we can now directly 
measure how many high-latitude Galactic
candidates selected using machine-learning classifiers
have been confirmed either as  pulsars or globular
clusters. Out of 22 Galactic candidates 
above $10\sigma$ significance picked by the 
classifiers in the 1FGL, 
18 have been confirmed as pulsars to date. This   
implies that the accuracy achieved by the classifiers exceeds 80\%
at that significance level.  In Figure 4 we show the
number of 3FGL Galactic candidates  at $|b| \geq 20\degree$  as a function of
flux. For comparison, 1FGL and 2FGL candidates using the
same machine-learning
classifiers are also shown. We can clearly see the tremendous
progress in
pulsar discoveries since the release of the
1FGL based on 11 months of \fermilat data.

\section{Prediction Results} 
Next we applied the classifier models to the entire 3FGL 
unassociated sample. 
In order to focus on sources outside the Galactic plane, 
we excluded unassociated 3FGL 
sources within 5\degree\ of the 
plane. This cut
leaves us with a starting list of 675 unassociated objects from a
total of 1010. 
Since we are only interested in high-quality predictions, 
we further impose the condition that
in order to retain a Galactic candidate 
both classifiers must agree (probability $P_{\rm Galactic} \geq 0.5$ that the object
is Galactic in both classifiers).  Finally, we only report
predictions for 3FGL objects with a detection
significance $\geq 10\sigma$.

Based on these criteria, 
we find a total of 34 high-latitude Galactic candidates with energy
flux $\gtrsim 4.0 \times10^{-12}$ erg
cm$^{-2}$ s$^{-1}$ between 100 MeV and 100 GeV
at $|b| \geq 5\degree$. Table 3 lists these candidates. 
To check for consistency, we compared our list of candidates
with results from spectral fitting of 3FGL unassociated sources
\citep{bertoni}, as well as with pulsar predictions using 
a combination of Random Forest and Logistic Regression \citep{saz}.
The list of
Galactic candidates at high Galactic latitude is
in good agreement with both of these works.

\bigskip
\section{Interpretations}
Armed with these 34 potential high-latitude Galactic
objects, we can now place the results in context of
known and hypothesized gamma-ray source classes.   

\subsection{Globular Clusters/Dwarf Galaxies: Optical Search}
One possibility is that some of these newly discovered objects are either
low-luminosity globular clusters  \citep{koposov,minniti}
or ultra-faint dwarf galaxies. In order to examine this possibility 
we have visually inspected Digitized Sky Survey (DSS) 
and Sloan Digital Sky Survey (SDSS) images \citep{ahn}. 
We confined our visual inspection to the area enclosed within 
the 3FGL 95\% uncertainty ellipses of the 34 candidates. Our
search mainly focused on visually diffuse and extended objects in
the optical,
but none were found at least 1 magnitude above the DSS optical limit.
It is important to note that the visual
approach is limited to objects that can easily standout in the images
but it is severely hampered when trying to reach ultra-faint
dwarf galaxies as the ones detected by SDSS \citep{willman}
and DES \citep{bechtol}.

For six optical fields with reliable multi-band SDSS photometry  
(3FGL J0318.1+0252, 3FGL J1120.6+0713,
3FGL J1225.9+2953, 3FGL J1625.1--0021, 3FGL J2103.7--1113,
3FGL J2212.5+0703),
we systematically searched for
unusual concentrations of RR Lyrae.  
RR Lyrae stars are excellent distance indicators, as well as superb tracers
of stellar substructures away from the Galactic plane
 \citep{mateo,alcock}. 
In order to pinpoint RR Lyrae star candidates within
the 3FGL 95\% uncertainty 
ellipses,  we adopted a simple SDSS color cut scheme based on 
Equations 1--4 derived by \citet{sesar}. 

We find no anomalous excess of RR Lyrae stars 
in any of the 6 fields with significance greater than $5\sigma$.
It is quite possible that 
fainter stellar structures with very few members could be present
below the optical detection limits \citep{willman,bechtol}.

\subsection{Pulsars: Comparison with Population Synthesis Models}
Given the ultra-fast pace of pulsar discoveries with {\it Fermi},
it is not surprising that 
at least five of the Galactic candidates identified here
have been already confirmed as pulsars since the 3FGL release (see Table 3).
In the 3FGL there are 89 pulsars listed
at $|b| \geq 5\degree$  with
energy flux larger than 
$\gtrsim 2.3 \times10^{-12}$ erg
cm$^{-2}$ s$^{-1}$ between 100 MeV and 100 GeV.
Of the 89 known high-latitude gamma-ray pulsars, 77 are MSPs
with spin period $P$ (s) and spin-down rate $\dot{P}$
satisfying log$_{10} \dot{P}$ + 19.5 + 2.5 log$_{10}$ P $<$ 0 \citep{3fgl}.
An addition of 34 pulsars  
would bring the total number of \fermilat detected pulsars at high
latitude to 123.

In order to compare this number to theoretical predictions from
pulsar simulations, 
we performed a new population synthesis of MSPs
\citep{gonthier1,gonthier2,gonthier3} in
a similar fashion as in the study of \citet{story}.  However, we have
used improved pulsar spin-down formalism, empirical gamma-ray luminosity, and
beam geometries of the high-energy emission models of the high-altitude
Slot Gap (two pole caustic) \citep{muslimova,muslimovb} (TPC), of the
Outer Gap \citep{chenga,chengb,zhang} (OG), and of the
pair-starved polar cap \citep{muslimovb,muslimovc} (PSPC).

In addition
to the radio beam geometry used in \citet{story}, we implemented a
radio aligned model (ALTPC) in which the radio emission has the same geometry
as the high-energy TPC model.  The characteristics of thirteen radio surveys
provided the detection thresholds for simulated radio pulsars, and the
{\it Fermi} point source threshold map in the 2FPC was scaled for various
observing periods to determine the detection of simulated gamma-ray
pulsars.

The 
models predict that \fermilat should detect anywhere from 100 to 126 MSPs
in four years with an
energy flux larger than $2.3\times10^{-12}$ erg
cm$^{-2}$ s$^{-1}$  at
$|b| \geq 5\degree$ (Table 4).
Assuming that 85\% of
the 34  Galactic candidates are MSPs, as is observed in the 3FGL catalog at high
Galactic latitude, then the total number of 
3FGL MSPs at $|b| \geq 5\degree$ would be $\approx 106$ (see Table 4).
Therefore, the simulated pulsar population
is in excellent agreement with the projected 
number of 3FGL MSP detections. In fact, depending on the pulsar
emission model there is 
some wiggle
room ($\approx$4--20) for additional 3FGL MSP discoveries at
high Galactic latitude. Below $10\sigma$ and above $4\sigma$ significance,
the machine-learning classifiers find an additional 33
Galactic candidates that could easily cover this difference.
The
locations of the 34 high-significance Galactic candidates and known 3FGL
pulsars
at $|b| \geq 5\degree$ are
shown in Figure 5.

\subsection{Dark Matter Subhalos: Comparison with Aquarius and
  Via Lactea II Numerical Simulations}
Taking advantage of the predictions 
in \citet{bertoni}, we can also directly compare the number of Galactic
candidates
at $|b| \geq 20\degree$ to the expected number of nearby subhalos detectable
by \fermilat after four years of observations.
Because our classifiers rely 
on the 3FGL variables, we are only sensitive to dark matter annihilation in
the 100 MeV--300 GeV energy range.
It is important to note that no significantly exponentially curved candidates
at energies $\gtrsim 300$ GeV have been found in any of the hard
{\it Fermi}-LAT 
source catalogs
\citep{2fhl}.
As in the original approach outlined in \citet{berlin},
the subhalo predictions in \citet{bertoni}
rely on the distribution of hundreds of thousands of subhalos that were
simulated in 6 ultra-highly resolved Milky-Way sized halos as part of  
the Aquarius Project \citep{springel}.

\citet{bertoni} calculated
the number of detectable Aquarius subhalos as a function of
\fermilat flux and Galactic latitude. Recently, \citet{djoeke} used results
of the Via Lactea II simulation \citep{diemand} 
scaled to the Planck 2015 cosmological
parameters and found a slightly smaller number of detectable subhalos,
but still consistent with the Aquarius results
considering the range of assumptions involved.  
For our comparison of high-latitude candidates with dark matter predictions,
we removed recently detected
pulsars from Table 3. This brings down
the number of $|b| \geq 20\degree$ subhalo candidates from 17 to 14.

Figure 6 shows the upper limits on $\langle\sigma v\rangle$
for dark matter masses $m_{\chi}$ between 30 GeV and 10 TeV annihilating into
a bottom quark 
($b\bar{b}$) final states based on the detection of 14 potential
subhalo candidates at 
$|b| \geq 20\degree$. Note that these limits would only be slightly
better if one includes the role of halo substructure to boost
the subhalo annihilation flux 
\citep{sanchez,moline}. More precisely, following
\citet{moline}, we estimate at most
a factor $\sim$ 10\% stronger limits when boosting the annihilation
signal in the mass range 
$10^{4}$--$10^{7}$ $M_{\odot}$. 
For comparison, we also show 
constraints if eventually no dark matter subhalos 
turn up in the 3FGL from the Aquarius \citep{bertoni} and 
Via Lactea II simulations \citep{djoeke} respectively.
While the current limit is not constraining enough to rule out
the canonical thermal relic cross
section at energies below 100 GeV \citep{steigman}, the curve is starting
to approach competitive values. Indeed,
further associations of some of these candidates with conventional
astrophysical sources may only lead to more stringent limits. 

As noted by \citet{bertoni}, there are
significant uncertainties in the number of predicted
dark matter subhalos that could shift these dark matter
limits by a factor of a few. 
The point remains that \fermilat might have detected several dark matter
subhalos by now, even under the most pessimistic assumptions. 
The absence of an overwhelming number of subhalo
candidates provides
complementary verification of more robust annihilating dark matter limits
inferred from dwarf galaxies \citep{geringer,dwarfs} 
and the Galactic center \citep{abazajian,gordon,calore2,daylan}.

\section{Conclusions and Outlook}
We find that the set of variables provided in the \fermilat catalogs
have the ability to effectively predict gamma-ray source classes
in the 3FGL dataset.
After careful examination of various Galactic demographics,
we find that the 34 additional high-latitude
Galactic candidates predicted using machine-learning
classifiers can be
accommodated by existing pulsar population synthesis models
without the need to introduce undiscovered globular clusters, 
dark matter subhalos, or gamma-ray emitting ultra-faint
dwarf galaxies. On the other hand, if these objects were produced by
annihilating dark matter, the upper limits on the annihilation cross
section are
starting to approach values
at or below the canonical thermal cross section for energies 
$\lesssim100$ GeV.

\begin{figure}
\epsscale{1.0}\plotone{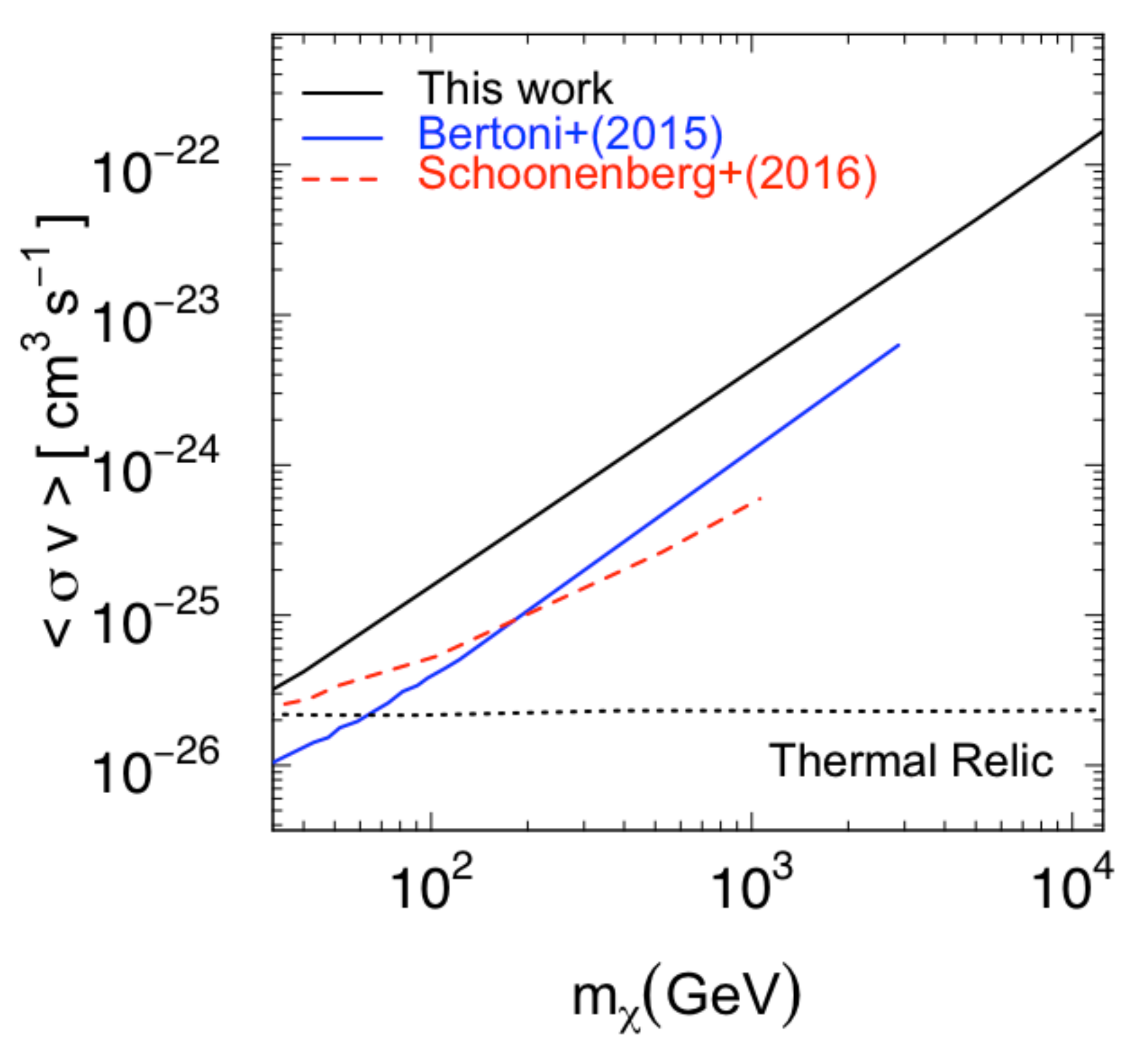}
\caption{Upper limits on the dark matter
  annihilation cross section for the 
$b\bar{b}$ channel assuming 14 subhalo
  candidates at $|b| \geq 20\degree$ (black solid line).
  The dashed red line 
is an upper limit derived from
the Via Lactea II simulation when zero 3FGL subhalos are adopted
\citep{djoeke}. The blue line corresponds to the constraint for
zero 3FGL subhalo candidates using the Aquarius simulation instead
\citep{bertoni}.
The horizontal dotted line marks the canonical thermal 
relic cross section \citep{steigman}.
}
\end{figure}

The discovery of radio and gamma-ray
pulsations will be crucial to address
the spectral
degeneracy between dark matter annihilation and pulsar emission.
However, blind searches will face greater obstacles in noisy
MSPs and fainter gamma-ray sources as \fermi continues operations.
Table 4 shows projected discoveries of MSPs for
10 years of \fermilat data taking.  
The most promising follow-up 
strategy to break these degeneracies will rest on our ability to detect  
pulsations going from the brightest to the faintest
Galactic candidates. Some of these searches for the most
elusive gamma-ray pulsars are being conducted by the distributed volunteer
computing sources,
$Einstein@Home$ \citep{pletsch}. New discoveries will require 
even larger computing resources and new search strategies.

Optical, ultraviolet and X-ray searches 
for binary objects with
temporal variability could also enhance the chances for finding millisecond
pulsars \citep{romani,bogdanov}. Incidentally, the addition
of new MSPs will also bring us closer to the detection of 
nanohertz gravitational waves based on pulsar-timing arrays 
\citep{taylor}. 
Should additional high-latitude Galactic candidates be confirmed as pulsars, 
new swaths of annihilation cross sections will
be disfavored by direct comparison with statistics from cosmological
numerical simulations of Milky Way-like galaxies.
Therefore, subhalo searches represent
a powerful complementary method to
existing probes of dark matter annihilation.

Clearly, there ought to be dedicated multiwavelength campaigns
to map the error ellipses of high-latitude Galactic
candidates for which no radio/gamma-ray pulsations are found. Finally,
the improvements
in position and photon flux afforded by  Pass 8 analysis \citep{atwood}
should further enhance machine-learning predictions in the future
\fermilat Fourth Source Catalog (4FGL).

\vspace {1cm}
{\it Acknowledgments.}
This research was supported by a senior 
appointment to the NASA Postdoctoral Program
at the Goddard Space Flight Center, administered by 
Universities Space Research Association through a contract with NASA.
MASC is a Wenner-Gren Fellow and acknowledges the support of the
Wenner-Gren Foundations to develop his research.
We thank Seth Digel for helpful suggestions.

\begin{deluxetable}{cccccc}
  \tablecaption{Machine-learning Galactic candidates among 3FGL unassociated sources at  $|b| \geq 5\degree$.}

\tablenum{3}

\tablehead{\colhead{Source Name} & \colhead{$l$} & \colhead{$b$} & \colhead{$P_{\rm Galactic}$(RF)} & \colhead{$P_{\rm Galactic}$(XGBoost)} & \colhead{ID or Assoc.}\\ 
\colhead{} & \colhead{(degrees)} & \colhead{(degrees)} & \colhead{} & \colhead{} & \colhead{}} 

\startdata
3FGL J0212.1+5320  & 134.93   & -7.65       & 1.00 & 1.00 \\
3FGL J0238.0+5237  & 138.85   & -6.92       & 0.94 & 0.85 &\\
3FGL J0312.1--0921 & 191.51   & -52.36      & 0.95 & 0.94 &\\
3FGL J0318.1+0252  & 178.45   & -43.64      & 1.00 & 1.00 &\\
3FGL J0336.1+7500  & 133.11   & 15.50       & 0.98 & 1.00 &\\
3FGL J0523.3--2528 & 228.20   & -29.83      & 0.89 & 1.00 &\\
3FGL J0545.6+6019 &   152.4964 &    15.7493 & 0.67 & 0.89 &\\    
3FGL J0758.6--1451 &   233.9599 &     7.5619 & 0.90 & 0.83 &\\
3FGL J0802.3--5610 &   269.9308 &   -13.1755 & 0.72 & 0.90 &\\
3FGL J0838.8--2829 &   250.6050 &     7.8008 & 0.97 & 1.00 &\\
3FGL J0933.9--6232 &   282.2351 &    -7.8937 & 1.00 & 1.00 &\\
3FGL J0953.7--1510 &   251.9380 &    29.6055 & 0.96 & 0.98 &\\
3FGL J0954.8--3948 &   269.8445 &    11.4575 & 0.61 & 0.52 &\\
3FGL J1035.7--6720 &   290.3908 &    -7.8284 & 1.00 & 1.00 &  psr \citep{camilo}\\
3FGL J1119.9--2204 &   276.4696 &    36.0588 & 0.97 & 1.00 &\\
3FGL J1120.6+0713 &   251.5322 &    60.6852 & 0.85 & 0.73 &\\
3FGL J1225.9+2953 &   185.1521 &    83.7648 & 0.94 & 1.00 &\\
3FGL J1539.2--3324 &   338.7592 &    17.5342 & 0.90 & 1.00 &\\
3FGL J1544.6--1125 &   356.2111 &    32.9844 & 0.63 & 0.73 & psr \citep{bogdanov}\\
3FGL J1557.0--4225 &   335.6413 &     8.3622 & 0.50 & 0.57 &\\
3FGL J1624.2--4041 &   340.5718 &     6.1421 & 1.00 & 1.00 &
PSR ($Einstein@Home$\tablenotemark{a})\\
3FGL J1625.1--0021 &    13.8808 &    31.8378 & 1.00 & 1.00 &\\
3FGL J1653.6--0158 &    16.6181 &    24.9246 & 0.98 & 1.00 &\\
3FGL J1702.8--5656 &   332.3978 &    -9.2447 & 0.62 & 0.84 &\\
3FGL J1744.1--7619 &   317.1046 &   -22.4711 & 0.99 & 1.00 & psr \citep{camilo}\\
3FGL J1753.6--4447 &   347.0854 &    -9.4164 & 0.98 & 0.99 &\\
3FGL J1946.4--5403 &   343.8883 &   -29.5630 & 1.00 & 0.99 & PSR \citep{ray}\\
3FGL J2039.6--5618 &   341.2312 &   -37.1551 & 0.98 & 1.00 &\\
3FGL J2103.7--1113 &    37.8579 &   -34.4231 & 0.93 & 0.87 &\\
3FGL J2112.5--3044 &    14.8984 &   -42.4487 & 0.99 & 0.99 &\\
3FGL J2117.6+3725 &    82.7982 &    -8.2737 & 0.81 & 0.90 &\\
3FGL J2133.0--6433 &   328.7390 &   -41.2683 &  0.93 & 1.00 &\\
3FGL J2212.5+0703 &    68.7429 &   -38.5650 & 0.92 & 0.99 &\\
3FGL J2233.1+6542 &   109.3427 &     6.5614 &  0.74 & 0.95 &\\
\enddata
\tablenotetext{a}{http://www.einsteinathome.org/}
\end{deluxetable} 

\begin{deluxetable}{ccccccc}
  \tablecaption{Summary of Population Synthesis of MSPs at  $|b| \geq 5\degree$.}

\tablenum{4}

\tablehead{\multicolumn{1}{c}{Catalog} & \multicolumn{1}{c}{Period} &
  \multicolumn{1}{c}{Detected MSPs + (Galactic Candidates)\tablenotemark{a}} & \multicolumn{4}{c}{Simulated}\\ 
  \colhead{} & \colhead{} & \colhead{} & \colhead{TPC} & \colhead{OG} & \colhead{AITPC} & \colhead{PSPC}} 
\startdata
3FGL  & 4 years  & 77 + (29) & 110 & 126 & 104 & 100\\
  & 5 years &  & 122 & 138 & 117 & 110\\
  & 10 years &  & 160 & 173 & 167 &   145\\
\enddata
\tablenotetext{a}{Number listed in parentheses corresponds to
  85\% of Galactic candidates,
  as in the observed percentage of MSPs in the 3FGL pulsar list
at high Galactic latitude.}
\end{deluxetable}

\end{document}